\def\be{\begin{equation}}
\def\ee{\end{equation}}
\def\ba{\begin{array}}
\def\ea{\end{array}}

\def\Rb{{I\!\! R}}

\def\Cb{\ \hbox{\vrule width 0.6pt height 5pt depth 0pt
		      \hskip -3.2 pt} C}
\documentstyle[12pt]{article}
\begin{document}
\parskip=4pt
\parindent=18pt
\baselineskip=22pt
\setcounter{page}{1}
\centerline{\Large\bf A Remark on the Optimal Cloning of An $N$-Level}
\centerline{\Large\bf Quantum System}
\vspace{6ex}
\begin{center}
{\large  Sergio Albeverio}\footnote {SFB 256; SFB  237; BiBoS;
CERFIM (Locarno); Acc.Arch., USI (Mendrisio)}~~~ and ~~~
{\large  Shao-Ming Fei}\footnote{Institute of Physics,
Chinese Academy of Science, Beijing.}
\end{center}
\begin{center}
Institut f\"ur Angewandte Mathematik, Universit\"at Bonn,
D-53115 Bonn\\
Fakult\"at f\"ur Mathematik, Ruhr-Universit\"at Bochum
D-44780 Bochum\\
\end{center}
\vskip 1 true cm
\parindent=18pt
\parskip=6pt
\begin{center}
\begin{minipage}{5in}
\vspace{3ex}
\centerline{\large Abstract}
\vspace{4ex}
We study quantum cloning machines (QCM) that act on
an unknown $N$-level quantum state and make $M$ copies.
We give a formula for the maximum of the fidelity of
cloning and exhibit the unitary transformations that
realize this optimal fidelity. We also extend the results to treat the
case of $M$ copies from
$N^\prime$ ($M>N^\prime$) identical $N$-level quantum systems.
\bigskip
\medskip
\bigskip
\medskip


\end{minipage}
\end{center}

\newpage

A major difference between cloning of classical and quantum information
is that whereas the copies of the classical information can be made perfect 
the copies of the quantum information are always imperfect, by the very 
principles of quantum theory. More precisely 
the classical information about the input
state is available through measurement and can be cloned by making a
measurement on the input state and using the result of the measurement
to make an arbitrary number $M$ of identical copies.
In contrast, from the very superposition principle of quantum
mechanics, an unknown quantum state can not be perfectly copied
(see e.g., \cite{Wootters,Barnum}). Consider an $N$-level quantum state. Let
$\vert i>$, $i=1,...,N$, be the basis vectors spanning
the Hilbert space of quantum states. A quantum state is a complex linear
combination of the bases vectors. For an arbitrary (unknown) input
state $\vert\Phi>$ and another given initial state $\vert\Phi>_0$ of the
blank copy, the no-cloning theorem (see e.g. \cite{Wootters}) says that
there is no unitary transformation $U$ of two quantum states
$\vert\Phi>$ and $\vert\Phi>_0$, such that
$U\vert\Phi>\vert\Phi>_0=\vert\Phi>\vert\Phi>$ (where $\vert\Phi>\vert\Phi>_0$
denotes the tensor product of $\vert\Phi>$ and $\vert\Phi>_0$).
Hence in cloning quantum states one has to drop the requirement that the
copies be perfect. The concept of quantum cloning 
machines (QCM) which act on an unknown quantum
state and make one, or more, imperfect copies of it has been introduced
\cite{Buzek}. To make the copies as good as possible
the transformations used by the quantum cloning machines should
be optimal, in the sense that they maximize the average fidelity between
the input and the output states.

The transformation that produces two copies of one
qubit state ($N=2$, i.e. a spin $\frac{1}{2}$ state),
with a fidelity independent of the state of the input qubit,
was first given in \cite{Buzek}. This transformation was shown to be
optimal \cite{Gisin,Bruss,Bruss1}. In \cite{Gisin} the transformations that
produce $M$ copies from $N^\prime$ ($M > N^\prime$) identical
states of qubits are also studied. Its optimality is generally proved
in \cite{Bruss1}. To clone entangled states of two or
more qubits, a transformation that produces two copies ($M=2$) from one
$N$-level quantum state was given in \cite{Buzekn}. 
In \cite{Werner} Werner studied optimal cloning of pure states for
turning a finite number of $N$-level quantum systems in the same unknown
state $\sigma$ into $M$ systems of the same kind, in an approximation of
the $M$-fold tensor product of the state $\sigma$.

In this article we study QCM that transform one $N$-level quantum state
into $M$ identical copies by using the representation of the algebra $SU(N)$,
a different approach from the one used in \cite{Werner}.
We first compute the maximum of the fidelity
for these QCM. Then we present the unitary transformation that realizes
this fidelity. More general QCM that produce $M$ copies from
$N^\prime$ identical $N$-level quantum systems ($M>N^\prime\geq 1$)
are also studied, by giving explicitly the unitary transformations and
the related fidelity. The optimal cloning of entangled states of many
qubits to $M$ copies are discussed. Our results recover the ones
obtained in \cite{Gisin,Buzekn,Werner} by taking different
values of $N$ and $M$.

For an $N$-level quantum state with basis $\vert j>$, $j=1,...,N$,
an input state is of the form
\be\label{psi}
\vert \Psi>=\sum_{j=1}^N\xi_j\vert j>,
\ee
where $(\xi_1,...\xi_N)$ is a point in the complex sphere
$S^{N-1}_{\Cb}$ with
$$
\ba{rcl}
\xi_1&=&e^{i\varphi_1}\sin\theta_{N-1}...\sin\theta_{2}\sin\theta_{1}\\[2mm]
\xi_2&=&e^{i\varphi_2}\sin\theta_{N-1}...\sin\theta_{2}\cos\theta_{1}\\[2mm]
...\\[2mm]
\xi_{N-1}&=&e^{i\varphi_{N-1}}\sin\theta_{N-1}\cos\theta_{N-2}\\[2mm]
\xi_{N}&=&e^{i\varphi_{N}}\cos\theta_{N-1}\\[2mm]
\ea
$$
where $i=\sqrt{-1}$, $0\leq\varphi_j<2\pi$, $0\leq\theta_j<\frac{\pi}{2}$,
$j=1,...,N$.

{\sf Theorem.} The maximal value of the fidelity $F$ for optimal cloning of an
$N$-level quantum state to $M$ copies is
\be\label{fmax}
F_{max}=\frac{2M+N-1}{M(N+1)}.
\ee

{\sf Proof}. Let $|R>$ be the initial state of the QCM and the $M-1$
blank copies. The most general action of the
QCM on $\vert \Psi>$ defined by (\ref{psi}) 
is given by a unitary operator $U$ such that
\begin{equation}\label{u}
\ba{rcl}
U|\Psi>|R>\equiv\vert\Psi_{out}>&=&
\displaystyle\sum_{i=1}^N\sum_{n_1,...,n_{N}=0}^M
\!\!\!\!\!\!\!\!^\prime~~~\xi_i
|n_1,...,n_N>|R_{i,n_1,...,n_N}>\\[5mm]
&\equiv&\displaystyle
\sum_{i=1}^N\sum_{\bf n=0}^M \!^\prime~\xi_i|{\bf n}>|R_{i,{\bf n}}>,
\ea
\end{equation}
where ${\bf n}$ denotes $n_1,...,n_N$, $\displaystyle\sum^{M}{^\prime}$
means to sum over the variables under the condition
that the sum of all the variables should be equal to $M$, i.e.
$\displaystyle\sum_{i=1}^N n_i=M$,
$|{\bf n}>=|n_1,...,n_N>$ is a completely symmetric (normalized) state
with $n_i$ quantum systems in the state $\vert i>$ (we have postulated that
the output of the QCM is completely symmetric, which does not
affect the conclusions, see the discussions in \cite{Gisin} for the case
$N=2$), $|R_{i,{\bf n}}>$ are unnormalized final states of the
additional $N$-level quantum systems contributing to the copies 
of the original quantum system.
By the unitarity of the evolution, the states $|R_{i,{\bf n}}>$
satisfy the relations
\begin{equation}\label{r}
\sum_{{\bf n}=0}^M \!^\prime~<R_{j^\prime,{\bf n}}|R_{j,{\bf n}}>
=\delta_{j^\prime,j}
\end{equation}
(with $<,>$ the scalar product in the Hilbert space).

As the output state is symmetric under permutations, the fidelity
of the copies is obtained by calculating the overlap of the reduced
density matrix of one copy, say the first, with the input state $|\Psi>$
and averaging over all input states:
\be\label{f}
\ba{rcl}
F &=& Tr\left[\displaystyle\sum_{i,i^\prime=1}^N
< \Psi_{out} |\xi_{i^\prime} |i^\prime><i|\xi^{*}_{i}
| \Psi_{out}>\right]\\[5mm]
&\equiv&\displaystyle\sum_{j,j^\prime=1}^N
\displaystyle\sum_{\bf n^\prime=0}^M \!^\prime~
\sum_{{\bf n}=0}^M \!^\prime~
<R_{j^\prime,{\bf n}^\prime}
|R_{j,{\bf n}}>A_{j^\prime,{\bf n}^\prime,j,{\bf n}},
\ea
\ee
where $A_{j^\prime,{\bf n}^\prime,j,{\bf n}}=
\displaystyle\sum_{i,i^\prime=1}^N
\int d{\rm \xi}\, \xi^{*}_{j^\prime}
\xi_{i^\prime} \xi^{*}_{i} \xi_{j}
Tr\left[<{\bf n^\prime}|i^\prime>< i|{\bf n}>\right]$ and
$d{\rm \xi}$ is the invariant measure on $S_{\Cb}^{N-1}$, i.e., in 
above spherical coordinates $\varphi=(\varphi_1,...,\varphi_N)$, 
$\theta=(\theta_1,...,\theta_{N-1})$:
$$
d{\rm \xi}\equiv d{\rm \xi}({\rm\varphi,\theta})=
\frac{(N-1)!}{2\pi^N}\prod_{r=1}^N d\varphi_r\prod_{k=1}^{N-1}\sin^{2k-1}
\theta_k\cos\theta_k d\theta_k.
$$

To get the maximum of $F$ as given by (\ref{f}), 
we impose the constraint of the
trace of eq. (\ref{r}), which gives the extrema of the fidelity
that is greater or equal to the one using the constraint of eq.
(\ref{r}). Using a corresponding
Lagrange multiplier $\lambda\in \Rb$, we have to extremize
$$
\ba{rcl}
F_\lambda&=&\displaystyle\sum_{j,j^\prime=1}^N
\displaystyle\sum_{\bf n^\prime=0}^M \!^\prime~
\sum_{{\bf n}=0}^M \!^\prime~
<R_{j^\prime,{\bf n}^\prime}|R_{j,{\bf n}}>
A_{j^\prime,{\bf n}^\prime,j,{\bf n}}\\[6mm]
&&-\lambda\left(<R_{j^\prime,{\bf n}^\prime}|R_{j,{\bf n}}>
\delta_{j^\prime,j}\delta_{{\bf n}^\prime,{\bf n}}-N\right),
\ea
$$
where $\delta_{{\bf n}^\prime,{\bf n}}=
\displaystyle\prod_{k=1}^N\delta_{n_k^\prime,n_k}$.

Varying with respect to the components of
$<R_{j^\prime,n_1^\prime,...,n_N^\prime}|$ one gets
\begin{equation}
\displaystyle\sum_{j=1}^N\displaystyle
\sum_{{\bf n}=0}^M \!^\prime~
\left(A_{j^\prime,{\bf n}^\prime,j,{\bf n}}
-\lambda\delta_{{\bf n}^\prime,{\bf n}}
\delta_{j^\prime,j}\right)|R_{j,{\bf n}}>=0.
\end{equation}
Hence the possible value of $\lambda$ are the eigenvalues of the matrix
$A_{j^\prime,{\bf n}^\prime,j,{\bf n}}$, with corresponding eigenvectors
$|R_{j,{\bf n}}>$. Multiplying the above equation
on the left by $<R_{j^\prime,{\bf n}^\prime}\vert$
and summing over $j^\prime,{\bf n}^\prime$ we have
\be\label{7}
\ba{l}
\displaystyle\sum_{j,j^\prime=1}^N
\displaystyle\sum_{\bf n^\prime=0}^M \!^\prime~
\sum_{{\bf n}=0}^M \!^\prime~
<R_{j^\prime,{\bf n}^\prime}
|R_{j,{\bf n}}>A_{j^\prime,{\bf n}^\prime,j,{\bf n}}\\[4mm]
~~~~=
\displaystyle\sum_{j,j^\prime=1}^N
\displaystyle\sum_{\bf n^\prime=0}^M \!^\prime~
\sum_{{\bf n}=0}^M \!^\prime~
<R_{j^\prime,{\bf n}^\prime}|R_{j,{\bf n}}>
\lambda\displaystyle\delta_{{\bf n}^\prime,{\bf n}}
\delta_{j^\prime,j} = N \lambda\,.
\ea
\ee
Comparing (\ref{f}) and (\ref{7}) we get
$F_\lambda=N\lambda$. Therefore the maximum of $F_\lambda$ is
proportional to the largest eigenvalue of the matrix
$A_{j^\prime,{\bf n}^\prime,j,{\bf n}}$.

By a straightforward calculation we have
$$
\ba{l}
<n_1^\prime,...,n_N^\prime|k><l|n_1,...,n_N>\\[4mm]
~~~=\left\{\ba{l}
\displaystyle\frac{1}{M}\sqrt{n_l(n_k+1)}
\delta{n^\prime_1,n_1}\,...\,\delta{n^\prime_l,n_l-1}\,...
\,\delta{n^\prime_k,n_k+1}\,...\,
\delta{n^\prime_N,n_N}~~~~~~~k\neq l\\[4mm]
\displaystyle\frac{n_l}{M}\delta_{{\bf n}^\prime,{\bf n}}~~~~~~~k=l
\ea
\right.
\ea
$$
and
$$
\int d{\rm \xi} \xi^{*}_{j^\prime}\xi_{i^\prime} \xi^{*}_{i} \xi_{j}=
\left\{\ba{ll}
\displaystyle\frac{2}{N(N+1)}~~~~~~~&i^\prime=j^\prime=i=j\\
\displaystyle\frac{1}{N(N+1)}&i^\prime=i\neq j^\prime=j
~~{\rm or}~~ i^\prime=j^\prime\neq i=j\\[4mm]
0&{\rm otherwise}
\ea
\right.
$$
Therefore
$$
\ba{rcl}
A_{j^\prime,n_1^\prime,...,n_N^\prime,j,n_1,...,n_N}&=&
\displaystyle\displaystyle\sum_{i,i^\prime=1}^N
\int d{\rm \xi} \xi^{*}_{j^\prime}\xi_{i^\prime} \xi^{*}_{i} \xi_{j}
\left[\displaystyle\frac{n_l}{M}\delta_{i^\prime,i}+\displaystyle\frac{1}{M}
\sqrt{n_l(n_k+1)}\vert_{i\neq i^\prime}\right]\\[3mm]
&=&\displaystyle\frac{1}{MN(N+1)}\left((m+n_j)\delta_{jj^\prime}
+\sqrt{n_j(n_{j^\prime}+1)}\vert_{j\neq j^\prime}\right).
\ea
$$
For any given $n_1,...,n_N$, if we arrange the matrix indices
$(j,n_1,n_2,n_3,...,n_N)$ as
$(1,n_1,n_2,n_3,...,n_N)$, $(2,n_1,n_2+1,n_3,...,n_N)$,
$(3,n_1,n_2,n_3+1,...,n_N)$, ..., $(N,n_1,n_2,...,n_N+1)$, then
$A$ is a block diagonal matrix. The block matrix $B$ is given by
$$
\ba{rcl}
B&=&\displaystyle\frac{1}{MN(N+1)}\cdot\\
&&\left(\ba{ccccc}
M+n_1&\sqrt{(n_2+1)n_1}&\sqrt{(n_3+1)n_1}&...&\sqrt{(n_N+1)n_1}\\
\sqrt{n_1(n_2+1)}&M+n_2+1&\sqrt{(n_3+1)(n_2+1)}&...&\sqrt{(n_N+1)(n_2+1)}\\
\sqrt{n_1(n_3+1)}&\sqrt{(n_2+1)(n_3+1)}&M+n_3+1&...&\sqrt{(n_N+1)(n_3+1)}\\
\vdots&&&&\vdots\\
\sqrt{n_1(n_N+1)}&\sqrt{(n_2+1)(n_N+1)}&\sqrt{(n_3+1)(n_N+1)}&...&M+n_N+1
\ea
\right),
\ea
$$
where $n_1=M-\displaystyle\sum_{i=2}^N n_i$. Setting
$\lambda=\frac{\lambda^\prime}{MN(N+1)}$ for some
$\lambda^\prime\in\Rb$, we have
$$
\ba{rcl}
\left\vert B-\lambda\right\vert
&=&\displaystyle\frac{n_1}{MN(N+1)}\prod_{i=2}^N(n_i+1)\cdot\\[6mm]
&&\left\vert\ba{ccccc}
1+\displaystyle\frac{M-\lambda^\prime}{n_1}&1&1&...&1\\
1&1+\displaystyle\frac{M-\lambda^\prime}{n_2+1}&1&...&1\\
1&1&1+\displaystyle\frac{M-\lambda^\prime}{n_3+1}&...&1\\
\vdots&&&&\vdots\\
1&1&1&...&1+\displaystyle\frac{M-\lambda^\prime}{n_N+1}
\ea
\right\vert\\[22mm]
&=&\displaystyle\frac{(M-\lambda^\prime)^{N-1}
(\lambda^\prime-2M-N+1)}{MN(N+1)}.
\ea
$$
Therefore the largest eigenvalue of the matrix $A$ is $\lambda_{max}=
\frac{\lambda^\prime_{max}}{MN(N+1)}=\frac{2M+N-1}{MN(N+1)}$.
The maximum of the fidelity is then obtained as
$$
F_{max}=F_{\lambda_{max}}=N\lambda_{max}=\frac{2M+N-1}{M(N+1)}.
$$
\hfill $\rule{3mm}{3mm}$

From formula (\ref{fmax}) we see that
when $N=2$ the fidelity is reduced to the one in \cite{Gisin}.
For $M=2$, formula (\ref{fmax}) gives the fidelity obtained from the
transformations in \cite{Buzekn} where
optimality had been tested numerically.
For large $N$, the optimal fidelity is almost independent of the number
of quantum levels of the qubit being copied, in fact we have
$\displaystyle\lim_{N\to\infty}F_{max}=\frac{1}{M}$. We also see that the
fidelity of the copies decreases with $N$, tending to $\frac{1}{M}$
as $N$ goes to infinity.

In the following we give a unitary transformation that realizes
the optimal fidelity given above. For any given $\vert i>$, $i\in
1,...,N$, the optimal cloning $\vert i>$ to $M$ copies
is given by the following transformation:
$$
U_{1,M}\vert i>\otimes R=\displaystyle
\sum_{{\bf n}_i=0}^M\!\!^\prime{}^\prime~
\alpha_{{\bf n}_i}\vert {\bf n}>\otimes R_{{\bf n}_i},
$$
where ${\bf n}_i$ denotes $n_1,...,n_{i-1},n_{i+1},...,n_N$,
$\displaystyle\sum^{M}{}\!^\prime{}^\prime$
means to sum over the variables under the condition
that $m_i\equiv n_1+...+n_{i-1}+n_{i+1}+...+n_N\leq M$,
$n_i$ in ${\bf n}$ is given by $M-m_i$, $R_{{\bf n}_i}$ are
orthogonal normalized internal states of QCM, and
$$
\alpha_{{\bf n}_i}=\sqrt{M-m_i}\sqrt{\frac{N!(M-1)!}{(M+N-1)!}}.
$$

The fidelity given by the above copy machine is
$$
\ba{rcl}
F&=&\displaystyle\sum_{{\bf n}_i=0}^M\!\!^\prime{}^\prime~
\displaystyle\frac{M-m_i}{M}\alpha_{{\bf n}_i}^2\\[6mm]
&=&\displaystyle\frac{N!(M-1)!}{M(M+N-1)!}
\displaystyle\sum_{n_1=0}^M...
\displaystyle\sum_{n_{i-1}=0}^{M-n_{i-2}}
\displaystyle\sum_{n_{i+1}=0}^{M-n_{i-1}}...
\displaystyle\sum_{n_N=0}^{M-m_i}(M-m_i)^2\\[6mm]
&=&\displaystyle\frac{2M+N-1}{M(N+1)}=F_{max}.
\ea
$$
Therefore $F_{max}$ is truly the maximum of the fidelity that can be
realized by optimal cloning.
An optimal copy of an input state (\ref{psi}) can be obtained
by the following unitary transformation,
\be\label{1m}
U_{1,M}\vert \Psi>\otimes R=\displaystyle
\sum_{i=1}^N\sum_{{\bf n}_i=0}^M\!\!^\prime{}^\prime~\xi_i\,
\alpha_{{\bf n}_i}\vert {\bf n}>\otimes R_{{\bf n}_i}.
\ee

We have discussed optimal cloning of an $N$-level quantum state to $M$
copies. Formula (\ref{1m}) can be also applied for making $M$ copies
from a quantum register with $N^\prime$ qubits such that
$2^{N^\prime}=N$. In the following we consider a quantum cloning
machine that takes $N^\prime$ identical $N$-level quantum systems into
$M$ identical copies ($M>N^\prime$). Let $\vert N^\prime i>$,
$i=1,...,N$, denote the input state consisting of $N^\prime$ quantum states
all in the state $\vert i>$. The quantum cloning machine is described by
\be\label{unm}
U_{N^\prime,M}\vert N^\prime\,i>\otimes R=\displaystyle
\sum_{{\bf n}_i=0}^{M-N^\prime}\!\!^\prime{}^\prime~
\beta_{{\bf n}_i}\vert {\bf n}>\otimes R_{{\bf n}_i},
\ee
$$
\beta_{{\bf n}_i}=\sqrt{\frac{(M-m_i)!}{(M-N^\prime-m_i)!}}
\sqrt{\frac{(N^\prime+N-1)!(M-N^\prime)!}{N^\prime!(M+N-1)!}},
$$
here $\displaystyle\sum^{M-N^\prime}{}\!^\prime{}^\prime$
means to sum over the variables under the condition
that $m_i\leq M-N^\prime$ so that the number $m_i$ of errors in the copies
is smaller or equal to the number $M-N^\prime$ of additional $N$-level
qubits. The fidelity of each output qubit is
\be\label{fnm}
F_{N^\prime M}=\displaystyle\sum_{{\bf n}_i=0}^M\!\!^\prime{}^\prime~
\displaystyle\frac{M-m_i}{M}\beta_{{\bf n}_i}^2
=\displaystyle\frac{M+N^\prime(M+N-1)}{M(N+N^\prime)}.
\ee

The unitary transformation (\ref{unm}) is a
generalization of (\ref{1m}). Its optimality can be proved for small
integer values of $N^\prime$. We believe that the method used in proving the
optimality for the case $N=2$ can be used for a general proof of the
optimality of the transformation (\ref{unm}). In fact, the dimension of
the state space for
$N^\prime$ $N$-level systems is $N^{N^\prime}$. If we view the
$N^\prime$ $N$-level systems as one of the states belonging to an
$N^{N^\prime}$-level system, then (\ref{unm}) is just a special case of
(\ref{1m}). However, with $\tilde{F}_{max}$ defined as $F_{max}$ with
$N$ replaced by $N^{N^\prime}$,
$\displaystyle \tilde{F}_{max}=\frac{2M+N^{N^\prime}
-1}{M(N^{N^\prime}+1)}\leq F_{N^\prime M}$, which implies that the more
one learns about the quantum input state, the better one can make a copy
of it. The explicitly given unitary transformations (\ref{1m}) and
(\ref{unm}) can help in constructing quantum computational networks for
the kinds of cloning machine we described.


\begin{thebibliography}{99}

\bibitem{Wootters} {W.K. Wootters and W.H. Zurek},
	       {Nature} {\bf 299}, 802 (1982).

\bibitem{Barnum}  {H. Barnum, C.M. Caves, C.A. Fuchs, R. Jozsa, and
	       B. Schumacher}, {Phys. Rev. Lett.} {\bf 76}, 2818 (1996).

\bibitem{Buzek}
V. Bu\v{z}ek and M. Hillery, Phys. Rev. A {\bf 54}, 1844 (1996).

\bibitem{Gisin}  {N. Gisin and S. Massar},
	      { Phys. Rev. Lett.} {\bf 79}, 2153 (1997).

\bibitem{Bruss}  {D. Bru\ss, D.P. DiVincenzo, A.K. Ekert, C. Macchiavello,
and J. Smolin}, Phys. Rev. A {\bf 57}, 2368(1998).

\bibitem{Bruss1}  {D. Bru\ss, A.K. Ekert and C. Macchiavello},
Phys. Rev. Lett. {\bf 81}, 2598(1998).

\bibitem{Buzekn}  {V. Bu\v{z}ek and  M. Hillery}, { Phys. Rev. Lett.}
              {\bf 81}, 5003 (1998).

\bibitem{Werner} R.F. Werner, Phys. Rev. A {\bf 58}, 1827(1998).

\end{thebibliography}
\end{document}